\documentclass[journal]{IEEEtran}
\usepackage{cite}
\usepackage{bm,graphicx}\usepackage[cmex10]{amsmath}
\usepackage{wrapfig}
\usepackage{array}
\usepackage{amssymb,epsfig}

\usepackage{tikz}
\usepackage{balance}
\usetikzlibrary{shapes,arrows}

\usepackage{mdwmath}
\usepackage{mdwtab}
\usepackage{multirow}

\ifCLASSINFOpdf
\else
\fi

\begin{document}

\title{Binary Continuous Phase Modulations Robust to a Modulation Index Mismatch}

\author{Malek~Messai,~\IEEEmembership{Member,~IEEE,}
        Giulio~Colavolpe, ~\IEEEmembership{Senior Member,~IEEE,}
        Karine~Amis,~\IEEEmembership{Member,~IEEE,}
        and~Fr\'ed\'eric~Guilloud,~\IEEEmembership{Member,~IEEE,}
\IEEEcompsocitemizethanks{\IEEEcompsocthanksitem M. Messai, K. Amis and F. Guilloud are with the Signal and Communication Department of Telecom Bretagne-Institut Telecom, 
CNRS Lab-STICC (UNR 6285), Brest 29200, France (e-mail:{malek.messai, karine.amis, frederic.guilloud}@telecom-bretagne.eu)
\IEEEcompsocthanksitem G. Colavolpe is with Universit\`a di Parma,
Dipartimento di Ingegneria dell'Informazione, Viale G. P. Usberti, 181A, I-43124 Parma, Italy, e-mail: giulio@unipr.it.}}

\maketitle

\begin{abstract}
We consider binary continuous phase modulation (CPM) signals used in some recent low-cost and low-power consumption telecommunications standard. When these signals are generated  through a low-cost transmitter, the real modulation index can end up being quite different from the nominal value employed at the receiver and a significant performance degradation is observed, unless proper techniques for the estimation and compensation are employed. For this reason, we design new binary schemes with a much higher robustness. They are based on the concatenation of a suitable precoder with binary input and a ternary CPM format. The result is a family of CPM formats whose phase state is constrained to follow a specific evolution.
Two of these precoders are considered. We will discuss many aspects related to these schemes, such as the power spectral density, the spectral efficiency, simplified detection, the minimum distance, and the uncoded performance. The adopted precoders do not change the recursive nature of CPM schemes. So these schemes are still suited for serial concatenation, through a pseudo-random interleaver, with an outer channel encoder. 

\end{abstract}

\IEEEpeerreviewmaketitle
 
\section{Introduction}

 Continuous-phase modulation (CPMs) signals \cite{AnAuSu86} are very interesting modulation formats which combine a constant signal envelope and excellent spectral efficiency properties \cite{BaFeCo09}. In particular, the constant envelope makes these modulations insensitive to nonlinear distortions and thus very attractive for an employment in satellite communications and in low-cost and low-power consumption transmitter standards. An analog implementation of the CPM modulator allows to further reduce the transmitter cost, at the expense of possible variations of the CPM waveform parameters around their nominal values. In particular, the modulation index will vary since it depends on the not well calibrated gain of the employed voltage-controlled oscillator (VCO). As an example, in Bluetooth operating in Basic Rate (BR) and Low Energy (LE) modes, the modulation index is specified to be in the intervals $[0.28, 0.35]$ and $[0.45, 0.55]$, respectively~\cite{blu_stand10}. The interval of the modulation index for the Ultra Low Energy (ULE) mode of the Digital Enhanced Cordless Telecommunication (DECT) standard is $[0.35, 0.7]$ \cite{esti_2013}. In the Automatic Identification System (AIS), the modulation index is nominally equal to $0.5$ but due to the imperfections of the AIS equipments, a variation of  $\pm 10\%$ is typically admitted~\cite{BoMiPrLeCoTo12}. 
 
On the other hand, the optimal maximum a-posteriori (MAP) sequence or symbol detectors for CPMs described in the literature, and implemented through the Viterbi or the BCJR algorithm, respectively, require a perfect knowledge of the modulation index at the receiver.  When a modulation index mismatch is present, a significant performance degradation is observed.

One possible solution can be the adoption of a noncoherent detector (e.g., see ~\cite{OsLu74,CoRa99b,LaScEnHu02,LaScJa05} and references therein), due to its robustness to the phase uncertainty induced by the imperfect knowledge of the modulation index. As an example, in~\cite{OsLu74} noncoherent detection of continuous-phase frequency shift keying (CPFSK) signals is carried over a sliding window and decision is made only on the middle bit of this window. This allows to limit the accumulated phase error due to the modulation index error.

Another alternative can be represented by the adoption at the receiver of an algorithm for the estimation of the modulation index~\cite{XuZh13} coupled with the low-complexity algorithms described in~\cite{MaFrKa14} or in~\cite{ZaGeKi12} which properly compensate the estimated error on the modulation index by using a per-survivor processing. More recently, the very general problem of soft-input soft-output (SISO) detection of a binary CPM signal with an unknown modulation index transmitted over a channel with phase noise has been considered in~\cite{MeCoAmGu15}.

All these schemes, however, operate at the receiver side and no attempt to increase the intrinsic robustness of the generated signal is made. This problem is addressed here. In other words, we will define new binary formats for which the performance degradation is very limited even when there is a significant modulation index mismatch between the transmitter and the receiver. These new schemes are based on the  concatenation of a precoder with binary input and ternary output, and a ternary CPM scheme. The aim of the precoder is to constrain the evolution of the CPM phase state. Two precoders will be described and investigated in this paper. 
%The overall scheme resulting from the adoption of the first precoder has only two states, independently of the adopted modulation index. The second precoder generates an overall scheme with a number of states which depends on the modulation index denominator and will be introduced to avoid the presence of impulses in the power spectral density.
We will show the properties of the power spectral density of these schemes and also study the uncoded performance and the spectral efficiency, which provides a benchmark on the coded performance, as discussed later. Suboptimal detection will be also considered. These schemes preserve the recursive nature of CPM formats which makes them very attractive when serially concatenated, through a pseudo-random interleaver, with an outer channel encoder. In fact, it is well known that, when the inner modulator/encoder is recursive, an interleaver gain is observed~\cite{BeDiMoPo98b}.

The paper is organized as follows. In Section~\ref{se:signals} we will review binary and ternary CPM signals and their Laurent decomposition. The proposed schemes and the corresponding detectors are described in Section~\ref{se:schemes}. Section~\ref{sec:spectrum} sheds some light on the power spectral density of these schemes and describes how the spectral efficiency is defined and computed. The performance analysis in case of uncoded transmission is investigated in Section~\ref{sec:analysis}. Simulation results are reported in Section~\ref{sec:sim_res}, whereas conclusions are drawn in Section~\ref{se:concl}.

\section{CPM Signals and Laurent Decomposition}
\label{se:signals}
In this section, we will briefly review binary and ternary CPMs and their Laurent decompositions~\cite{La86,PeRi08} since they have relevance for the schemes proposed in this paper.

The complex envelope of a CPM signal can be expressed as~\cite{AnAuSu86}
\begin{equation}
s(t,{\bm a},h)=\sqrt{\frac{E_S}{T}} \exp \Big\{\jmath 2 \pi h \sum_n a_n q(t-nT) \Big\},   \label{e:bdb_sig}
\end{equation}
where $E_S$ is the energy per information symbol,
$T$ the symbol interval, $h$ the modulation index, the function $q(t)$ is the {\em phase  response}, and its derivative is the {\em frequency pulse}, assumed of duration $LT$ and integral 1/2. The information symbols ${\bm a}=\{a_n\}$, assumed independent, belong to the  alphabet $\{\pm 1\}$ in the case of binary CPMs and to the alphabet $\{0,\pm 2\}$ in the case of ternary CPMs.

The modulation index is usually written as~$h=r/p$ (where $r$ and $p$ are relatively prime integers).
In this case, it can be shown~\cite{Ri88} that the CPM signal in the generic time interval~$[nT,(n+1)T]$ is completely defined by the symbol~$a_{n}$, the \emph{correlative state}
$$
	\omega_{n}=(a_{n-1},a_{n-2},\dots,a_{n-L+1})
$$
and the \emph{phase state}~$\phi_{n-L}$, which takes on $p$ values and can be recursively defined as
\begin{equation}
	\phi_{n-L}=\left[\phi_{n-L-1}+\pi ha_{n-L}\right]_{2\pi}\;,\label{e:phase_state}
\end{equation}
where $[\cdot]_{2\pi}$ denotes the ``modulo $2\pi$'' operator. 
%In other words, we may express the complex envelope of a CPM signal as (\emph{Rimoldi decomposition})
%$$
%	s(t,\bm{a})=\sqrt{\frac{2E_{S}{T}}\sum_{n=0}^{N-1}s_{T}(t-nT;a_{n},\omega_{n})e^{j\phi_{n}}
%$$
%where $s_{T}(t-nT;a_{n},\omega_{n})$ is a slice of signal of length $T$ (with support in $[nT,(n+1)T]$) whose shape only depends on symbol $a_{n}$ and correlative state $\omega_{n}$ and is independent of the considered symbol interval.

Based on Laurent representation, the complex envelope of a binary or a ternary CPM signal may be exactly expressed as~\cite{La86,PeRi08}
\begin{equation}    
s(t,{\bm a},h)=\sum_{k=0}^{K-1}\sum_{n}\alpha_{k,n}   
p_{k}(t-nT) \label{e:Laurent} 
\end{equation} 
where $K$ represents the number of linearly-modulated components in the representation. It results to be $K=2^{(L-1)}$~\cite{La86} or $K=2\cdot 3^{(L-1)}$~\cite{PeRi08} for binary or ternary CPMs, respectively.
The expressions of pulses $\{p_{k}(t)\}$ as a function of $q(t)$ and $h$
and those of symbols $\{\alpha_{k,n}\}$ as a function of the information symbol
sequence $\{a_n\}$ and $h$ may be found in \cite{La86,PeRi08}. By truncating the summation in (\ref{e:Laurent}) considering only the first \mbox{$K < 2^{(L-1)}$} terms, we obtain an approximation of $s(t,{\bm a},h)$.

In the binary case, most of the signal power is concentrated in the first component, i.e., that associated with pulse $p_{0}(t)$, which is called \textit{principal component}~\cite{La86}.  As a consequence, the principal component may be used in (\ref{e:Laurent}) to attain a very good trade-off between approximation quality and number of signal components~\cite{CoRa97b,BaCo07b}. In this case, it holds
\begin{equation}
\alpha_{0,n} = \alpha_{0,n-1} e^{\jmath\pi h a_n}
 \label{eq:alpha}
\end{equation}
Symbols $\{\alpha_{0,n}\}$ take on $p$ values~\cite{La86} and it can be easily observed from (\ref{e:phase_state}) and (\ref{eq:alpha}) that $\alpha_{0,n}=e^{\jmath\phi_{n}}$.

In the ternary case, most of the signal power is concentrated in two principal components, corresponding to pulses $p_0(t)$ and $p_1(t)$. In this case, the corresponding symbols can be expressed as
\begin{align}
\alpha_{0,n} &=  e^{\jmath\pi h a_n}\alpha_{0,n-1}\\
\alpha_{1,n} &= \frac{1}{2} \left[e^{\jmath\pi h \gamma_{0,n}}+e^{\jmath\pi h \gamma_{1,n}}\right]\alpha_{0,n-1}
\end{align}
where $\gamma_{0,n}$ and $\gamma_{1,n}$ belong to the alphabet ${\pm 1}$ and are such that $a_n=\gamma_{0,n}+\gamma_{1,n}$~\cite{PeRi08}. In this ternary case, it is again $\alpha_{0,n}=e^{\jmath\phi_{n}}$.

In low cost transmitters, the value of the modulation index is often different from its nominal value which is instead assumed at the receiver. In the following, we will express the modulation index at the transmitter as $h=h_{rx}+h_e$, where $h_{rx}$ is the nominal value known at the receiver and $h_e$ accounts for the mismatch between transmitter and receiver and is assumed unknown. This mismatch has a catastrophic effect on the performance. As observed in~\cite{XuZh13}, the pulses of the principal components weakly depend on the value of the modulation index and thus on $h_e$. On the contrary, the effect of $h_e$ is cumulative in the phase state (\ref{e:phase_state}) and thus in $\alpha_{0,n}$.  
This observation motivates the schemes proposed in the next section.

We consider transmission over an additive white Gaussian noise (AWGN) channel. The complex envelope of the received signal thus reads
\begin{equation}
r(t) =   s(t,{\bm a},h_{rx}+h_e) + w(t)   \label{e:r(t)}
\end{equation}
where $w(t)$ is a complex-valued white Gaussian noise process with independent components, each with two-sided power spectral density $N_0/2$. In the following, we will denote by $\boldsymbol{r}$ a suitable vector of sufficient statistics extracted from the continuous-time received signal $r(t)$.

\section{Proposed Schemes and Corresponding Detectors}
\label{se:schemes}

\begin{figure}[!t]
\center
\includegraphics[width=6cm,height=3.5cm]{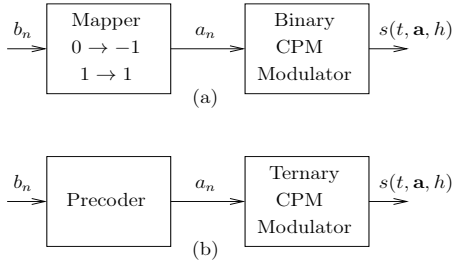}
\caption{Compared schemes. (a) Classical binary CPMs. (b) Proposed scheme.}
\label{fig:schemes}
\end{figure}

Our aim is to improve the robustness of classical binary CPM schemes to a modulation index mismatch. A classical binary CPM modulator can be represented as depicted in Fig.~\ref{fig:schemes}(a). Information bits $\{b_n\}$, belonging to the alphabet $\{0,1\}$ are first mapped into symbols $\{a_n\}$ belonging to the alphabet $\{\pm 1\}$, and then go at the input of a binary CPM modulator.
The schemes proposed in this paper are instead  based on the serial concatenation of an outer precoder, which receives at its input bits $\{b_n\}$ belonging to the alphabet $\{0,1\}$ and provides at its output ternary symbols $\{a_n\}$ belonging to the alphabet $\{0,\pm 2\}$, and an inner ternary CPM. This concatenation is shown in  Fig. \ref{fig:schemes}(b). It is reminiscent of uncoded shaped-offset quadrature phase-shift keying (SOQPSK)~\cite{PeRi08} although the precoder is designed there for different purposes, i.e., to  allow a simple (albeit suboptimal) symbol-by-symbol detection architecture and the ternary CPM scheme is generic here. 

We design the precoder to avoid the cumulative effect of $h_e$ on the phase state, still keeping its recursive definition. This latter property is important to have an interleaver gain when the proposed schemes are concatenated with an outer encoder through an interleaver. 

In order to explain the main ideas behind the proposed schemes, let us consider the first proposed precoder. It is simply the classical alternate mark inversion (AMI) precoder where bit $b_n=0$ is encoded as $a_n=0$,  whereas $b_n=1$ is encoded alternately as $a_n=2$ or $a_n=-2$. By denoting $\varphi= 2\pi h$ and assuming that the initial phase state is $\phi_0=0$, the arrival of a bit $b_n=0$ will leave the phase state in the previous value, i.e., $\phi_n=\phi_{n-1}$. On the contrary, the arrival of a bit $b_n=1$ will produce a change of phase state according to the following rule:
$$
\phi_{n}=\begin{cases}
\varphi & \mbox{if \ensuremath{\phi_{n-1}=0}}\\
0 & \mbox{if \ensuremath{\phi_{n-1}=\varphi}\,.}
\end{cases}
$$

\begin{figure}[!t]
\center
\includegraphics[width=8cm]{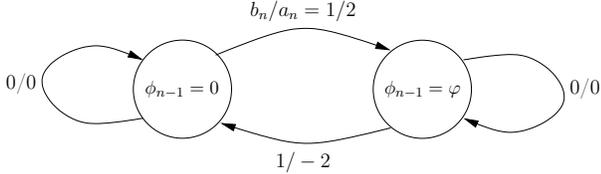}
\caption{State diagram of the overall scheme in the case of AMI precoder.}
\label{fig:AMIprecoder}
\end{figure}

We thus have the alternation of the phase states $0$ and $\varphi$ according to the state diagram shown in Fig.~\ref{fig:AMIprecoder}. In this state diagram, the values of $a_n$ are also shown. Thus, there is no accumulation effect and a possible error on $h$ has a very limited impact on the performance, as shown in Section~\ref{sec:sim_res}. 

\begin{figure}[!t]
\center
\includegraphics[width=9cm]{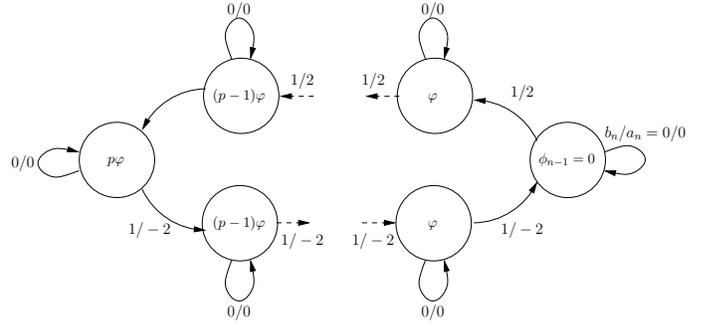}
\caption{State diagram of the overall scheme in the case of second proposed precoder.}
\label{fig:second_precoder}
\end{figure}

As we will see in Section~\ref{sec:spectrum}, unless $h=1/2,3/2,\dots$, the power spectral density of the signal resulting from the previous scheme has impulses at harmonics of the signaling rate $1/T$. They are present when $E\{\alpha_{0,n}\}=E\{e^{\jmath \phi_n}\}\neq 0$ (and this explains that when $h=1/2,3/2,\dots$, i.e., $\varphi=\pi,3\pi,\dots$, these impulses are not present).
These impulses can be used to help timing and frequency synchronization. In case they are considered as undesirable, another precoder can be used. This precoder must be defined such that all phase states 
$$
\Big\{0,\varphi,2\varphi,\dots, \big(p - 1\big)\varphi \Big\}
$$
%$$
%\Big\{-p \varphi, \big(-p + 2\big) \varphi,\dots,-2 \varphi,0,2 \varphi,\dots, \big(p - 2\big)\varphi,p \varphi \Big\}
%$$
%when $p$ is even and 
%$$
%\Big\{-(p-1)\varphi, -(p-3)\varphi,\dots,-2 \varphi,0,2 \varphi,\dots, (p-3)\varphi,(p-1)\varphi \Big\}
%$$
%when $p$ is odd, 
occur with the same probability. In this way, it will be $E\{\alpha_{0,n}\}=E\{e^{\jmath\phi_n}\}= 0$ since equally spaced discrete values on the unit circles are taken by the phase state.
One possible solution is the adoption of a precoder based on this simple rule: bit $b_n=0$ is again encoded as $a_n=0$, and thus leaves the CPM signal into the same state, whereas
bit $b_n=1$ is encoded as $a_n=2$ or $a_n=-2$. This time, however, the encoder provides at its output a block of $p$ symbols $a_n=2$ followed by a block of $p$ symbols $a_n=-2$, one after the other. Without loss of generality, let us assume that the initial phase state of the modulator is $\phi_0=0$.
The state of the overall scheme is not only related to the actual CPM phase state $\phi_{n-1}$ but also to the sign of the block of symbols we are transmitting. Starting from the phase state $\phi_0=0$, the first $p$ input bits 1 will drive the phase state to the state $p \varphi=0$ passing through the states $\varphi,2\varphi,\dots,(p-1)\varphi$. Then, the successive block of $p$ input bits 1 will take back the phase state to the initial value. Thus, these phase states are taken with  the same probability and 
$E\{\alpha_{0,n}\}=E\{e^{\jmath\phi_n}\}= 0$. The state diagram of the overall scheme is shown in Fig.~\ref{fig:second_precoder}, where states $\phi_{n-1}=0$ and $\phi_{n-1}=p\varphi$, which are the same phase state, have been split because conceptually different (the beginning and the ending state of a symbols block $a_n=2$).

As far as detection is concerned, we could implement the optimal receivers through a bank of suitable matched filters followed by a Viterbi~\cite{Fo73} or a BCJR~\cite{BaCoJeRa74} algorithm with a proper number of states.\footnote{The Viterbi and the BCJR algorithms are used for the implementation of the maximum a posteriori (MAP) \textit{sequence} and  \textit{symbol} detection criteria, respectively.} However, a  low-complexity suboptimal receiver with practically optimal performance can  be implemented through a bank of two filters matched to pulses $p_0(t)$ and $p_1(t)$, followed by a Viterbi or a BCJR algorithm with branch metrics~\cite{CoRa97b,BaCo07b}\footnote{Among many algorithms proposed in the literature for suboptimal detection of CPM signals (as an example, see the references in~\cite{CoRa97b,BaCo07b}), we here consider only those based on the Laurent decomposition since, thanks to the property mentioned before that pulses of the principal components weakly depend on $h_e$, they can employ fixed front end filters.}
\begin{align}
\lambda_n(b_n,\phi_{n-1})=&\,\Re\left[x_{0,n}\alpha^*_{0,n}+
x_{1,n}\alpha^*_{1,n}
\right]\nonumber \\
=&\,\Re\Big\{e^{-\jmath \phi_{n-1}}\Big[x_{0,n}e^{-\jmath\pi h a_n}\nonumber\\
&+x_{1,n}\Big(\frac{e^{-\jmath\pi h \gamma_{0,n}}+e^{-\jmath\pi h \gamma_{1,n}}}{2}\Big)\Big]
\Big\}
\end{align}
$x_{0,n}$ and $x_{1,n}$ being the outputs at time $nT$ of the two matched filters, having impulse response $p_0(-t)$ and $p_1(-t)$, respectively, $a_n$ is a function of $\phi_{n-1}$ and $b_n$, as shown in Fig.~\ref{fig:AMIprecoder} or in Fig.~\ref{fig:second_precoder}, and, as previously stated, $\gamma_{0,n}$ and $\gamma_{1,n}$ are such that $a_n=\gamma_{0,n}+\gamma_{1,n}$. These algorithms thus work on a trellis with only 2 states, independently of the value of $h$, in the case of the AMI precoder, whereas in the case of the second proposed encoder the number of states is $2p$. 

From a complexity point of view of a suboptimal detector, let us compare a classical binary CPM with the proposed precoders for a same modulation index and frequency pulse. For classical binary CPMs a single matched filter is required whereas in the case of the proposed schemes, two matched filters of the same length are employed. The front end complexity is thus doubled.

Let us now consider the computational complexity in each trellis section. For the proposed precoders, the computation of a single branch metric requires to compute two terms $x_{0,n}\alpha^*_{0,n}+
x_{1,n}\alpha^*_{1,n}$ instead of a single term $x_{0,n}\alpha^*_{0,n}$. So the complexity for each trellis branch is doubled. Regarding the number of branches in a single trellis section, which directly influences the receiver complexity and is given by the number of trellis states times the number of branches at the output of each state (2 in any case), this will be $p\times 2=2p$ in the case of a classical CPM, $2\cdot 2=4$ for the AMI precoder, and $2p\times 2=4p$ for the second precoder. As a consequence, we can conclude that in the case of the AMI precoder the computational complexity in each trellis section  is $4/p$ times that for a classical CPM, whereas for the second proposed precoder it is 4 times that for a classical CPM.

We can now explain why the proposed schemes are more robust in the case of modulation index mismatch, as an example with reference to the second precoder. Since the phase state starts from a certain value, evolves as a consequence of the arrival of $p$ input bits equal to one, and then comes back to that original value, the phase errors due to a modulation index mismatch do not accumulate and will be undone after $2p$ input bits equal to one. We can also expect the AMI precoder to be more robust than the second proposed precoder since the range of phase values, and thus the possible phase errors, is much more limited.

\section{Spectrum and Spectral Efficiency} 
\label{sec:spectrum}

In order to gain a deeper understanding of the effect of these precoders on the overall signal, we now consider the power spectral density (PSD) of the transmitted signal in our proposed schemes. Instead of considering the exact PSD, we will consider the PSD of the signal resulting from the approximation of the transmitted signal with its two principal components.\footnote{When $L=1$, this representation with only two components turns out to be exact, i.e., in (\ref{e:Laurent}) only two components are present in this case.} Hence, we approximate the transmitted signal as
\begin{equation}    
s(t,{\bm a},h)\simeq \overline{s}(t,{\bm a},h)=\!\!\sum_{n}\alpha_{0,n}   
p_{0}(t-nT)+\!\sum_{n}\alpha_{1,n}   
p_{1}(t-nT) \label{e:Laurent_app} 
\end{equation}
whose PSD can be expressed as 
\begin{align}
W_{\overline{s}}(f)=&\frac{W_{\alpha_0}(f)}{T}|P_0(f)|^2
+\frac{W_{\alpha_1}(f)}{T}|P_1(f)|^2\nonumber\\
&+\frac{2}{T}\Re\Big\{W_{\alpha_0,\alpha_1}(f)P_0(f)P^*_1(f)\Big\}
\label{e:PSD}
\end{align}
where $W_{\alpha_0}(f)$, $W_{\alpha_1}(f)$, and $W_{\alpha_0,\alpha_1}(f)$ are the Fourier transforms of the autocorrelations and crosscorrelations $R_{\alpha_0}(m)=E\{\alpha_{0,n+m}\alpha^*_{0,n}\}$, $R_{\alpha_1}(m)=E\{\alpha_{1,n+m}\alpha^*_{1,n}\}$, and $R_{\alpha_0,\alpha_1}(m)=E\{\alpha_{0,n+m}\alpha^*_{1,n}\}$, respectively, whereas $P_0(f)$ and $P_1(f)$ are the Fourier transforms of pulses $p_0(t)$ and $p_1(t)$, respectively.
The first two terms in (\ref{e:PSD}) represent the PSD of the two components whereas the remaining one takes into consideration the correlation between the two components.
It is straightforward to prove that, for the precoder in Fig.~\ref{fig:AMIprecoder} it is
\begin{equation}
 R_{\alpha_0}(m) = \left\{
      \begin{aligned}
      &  1, \quad \mbox{for $m=0$}  \\
      &  \frac{1}{2} (1+\cos 2 \pi h) , \quad \mbox{otherwise} \\
      \end{aligned}
    \right.
\end{equation}
\begin{equation}
 R_{\alpha_1}(m) =  \left\{
    \begin{aligned}
   & \frac{1}{2}+\frac{1}{2}\cos^2\pi h , \quad  \mbox{for $m=0$} \\
   & \frac{1}{4}+\frac{3}{4}\cos^2\pi h , \quad \mbox{for $m=\pm 1$}\\ 
   & \frac{\cos^2\pi h}{2}(1+\cos2 \pi h), \quad  \mbox{otherwise} \\
    \end{aligned}
  \right.
\end{equation}
and
\begin{equation}
 R_{\alpha_0,\alpha_1}(m) =  \left\{
    \begin{aligned}
   &  \cos\pi h, \quad  \mbox{for $m=0,-1$}   \\
   & \frac{3}{4} \cos \pi h + \frac{1}{4}\cos \pi h\cos 2 \pi h     , \quad\mbox{otherwise.} \\
    \end{aligned}
  \right.
\end{equation}

%it is \textbf{(Malek, please check if this formula is correct. Do you have some simpler closed forrm expression for $W_{\overline{s}}(f))$ in the case of both precoders?}
%\begin{equation}
%W_{\overline{s}}(f)=\frac{1}{T}\Big[
%C(f)-|\mu(f)|^2
%\Big]+\frac{1}{T^2}|\mu(f)|^2\sum_{\ell=-\infty}^\infty\delta\Big(f-\frac{\ell}{T}\Big)
%\label{e:PSD_precoder1}
%\end{equation}
%where $\delta(\cdot)$ is the Dirac delta, and
%\begin{aligned}
%C(f)=&|P_0(f)|^2+\frac{1}{2}|P_1(f)|^2\Big(1+\cos^2\pi h\Big)\nonumber \\
%&+2\Re[P_0(f)P^*_1(f)]\cos(\pi h)\nonumber\\
%&+\frac{1}{2}\Re\Big[P_1(f)P^*_0(f)e^{-\jmath 2\pi f T}\Big]\cos\pi h \sin^2(\pi h)\nonumber\\
%&+\frac{1}{2}|P_1(f)|^2\cos^2\pi h \sin^2(\pi h) \cos(2\pi f T)\nonumber\\
%\mu(f)=&P_0(f)\cos(\pi h)\frac{1}{2}P_1(f)\Big[1+\cos^2\pi h\Big]\,.
%\end{aligned}
As mentioned in the previous section,  this PSD presents some impulses, generated by the fact that, unless $h$ is such that $\cos \pi h=0$, it is,
according to the autocorrelation function properties~\cite{PaPi02},
\begin{align}\label{e:r_a_n}
\lim_{m\rightarrow \infty}R_{\alpha_0}(m)&=|E\{\alpha_{0,n}\}|^2\neq 0 \nonumber\\
\lim_{m\rightarrow \infty}R_{\alpha_1}(m)&=|E\{\alpha_{1,n}\}|^2\neq 0\nonumber\\
\lim_{m\rightarrow \infty}R_{\alpha_0,\alpha_1}(m)&=E\{\alpha_{0,n}\}E\{\alpha^*_{1,n}\}\neq 0\,.\nonumber 
\end{align}
However, it is sufficient to adopt a precoder such that in Fig.~\ref{fig:second_precoder} that makes $E\{\alpha_{0,n}\}=0$ to avoid the presence of such impulses. In fact, for ternary CPMs, by considering all symbols $\{\alpha_{k,n}\}$ (not
only those corresponding to principal components) it can be easily shown that they can be expressed as~\cite{PeRi08}
\[
\alpha_{k,n}=\alpha_{0,n-\ell}f(a_{n},\dots,a_{n-\ell+1})
\]
for a suitable $\ell$, where $f(\cdot)$ is a suitable (nonlinear) function. Since $\alpha_{0,n-\ell}$ is independent of future symbols $a_{n},\dots,a_{n-\ell+1}$, $E\{\alpha_{0,n}\}=0$ is a sufficient condition for having $E\{\alpha_{k,n}\}=0,\,\forall k$. Thus we have 
\begin{align}
\lim_{m\rightarrow \infty}R_{\alpha_k}(m)&=|E\{\alpha_{k,n}\}|^2= 0 \nonumber\\
\lim_{m\rightarrow \infty}R_{\alpha_{k_1},\alpha_{k_2}}(m)&=E\{\alpha_{k_1,n}\}E\{\alpha^*_{k_2,n}\}= 0\nonumber
\end{align}
and no impulses will be present in the resulting PSD. The computation of the PSD when the precoder in Fig.~\ref{fig:second_precoder} is used is much more involved and is not reported here.

%As an example, in the case of a 1REC (in this case, being $L=1$, $s(t)$ and $\overline{s}(t)$ coincide) frequency pulse with $h=1/4$ the PSD are shown in the following figure.

%\textcolor{red}{The PSD in the case of a 1REC frequency pulse with $h=1/4$ are given in Fig.~\ref{fig:dsp} for illustration purpose. We observe that simulated and theoretical PSD  coincide in the AMI precoder and that they exhibit impulses contrary to the second precoder.} 
 
%\begin{figure}[!h]
%\center
%\includegraphics[width=9cm]{figs/dsp_1REC_matlab.eps} %,height=5cm
%\caption{PSD for a 1REC frequency pulse with $h=1/4$.}
%\label{fig:dsp}
%\end{figure}

The precoder has effect not only on the PSD but also on the performance of the overall resulting scheme. In order to gain a deeper understanding of the schemes resulting from the proposed concatenations, we also evaluated their spectral efficiency (SE). The spectral efficiency $\eta$ is defined as 
\begin{equation}
\eta=\frac{I}{BT}\,,\quad[\textrm{bit}/\textrm{s}/\textrm{Hz}]\label{eq:se}
\end{equation}
where $I$ is
the information rate, i.e., the amount of information transmitted per channel use, and $B$ is the bandwidth occupied by the transmitted signal. This normalization is required to capture the different bandwidth occupancy of different modulation formats. In other words,  by considering the spectral efficiency, i.e.,  the amount of information transmitted per unity of time and
per unity of bandwidth,  we are also considering the effect of the precoding on the bandwidth occupancy. Our aim is to understand if the proposed binary schemes have a SE comparable with that of the classical binary schemes for a same frequency pulse and modulation index, and thus a similar performance has to be expected when the proposed signals are employed in coded systems.

CPM bandwidth is theoretically infinite
because the PSD of a CPM signal has rigorously an infinite support.
Hence, we consider the traditional definition of bandwidth based on
the power concentration, that is the bandwidth that contains a given
fraction of the overall power. Being this fraction a parameter, we
choose to use the $99.9\%$ of the overall power. This definition
is consistent with systems where a limitation on the out-of-band power
exists. To compute this bandwidth, we need the CPM power spectral density,
which cannot be evaluated analytically in closed form but only numerically.\footnote{In order to avoid approximations, we didn't use (\ref{e:PSD}). Instead, we adopted the technique described in~\cite{AnAuSu86} for the computation of the PSD.} 
An alternative bandwidth definition could be adopted. In particular, we could refer to the definitions employed in~\cite{AnAuSu86,BaFeCo09}. This would change our considerations only from a quantitative point of view.

To compute the information rate $I$ in~(\ref{eq:se}), we can
use the simulation-based technique described in \cite{ArLoVoKaZe06},
which only requires the existence of an optimal MAP symbol detector
for the considered system.  The simulation-based method described
in \cite{ArLoVoKaZe06} allows to evaluate the achievable information
rate as 
\begin{equation}
I\!(\boldsymbol{a},\boldsymbol{r})=E\left\{ \!\log\frac{p(\boldsymbol{r}|\boldsymbol{a})}{p(\boldsymbol{r})}\!\right\}\,, \quad[\textrm{bit}/\textrm{ch.use}]\label{eq:AIR}
\end{equation}
where the probability density functions $p(\boldsymbol{r}|\boldsymbol{a})$
and $p(\boldsymbol{r})$ 
can be evaluated recursively through the forward recursion of the optimal
MAP symbol detection algorithm, thus in the absence of modulation index mismatch.
This receiver can assure communication with arbitrarily
small non-zero error probability when the transmission rate at the
CPM modulator input does not exceed $I(\boldsymbol{a},\boldsymbol{r})$
bits per channel use, provided that a suitable channel code is adopted.

\section{Uncoded Performance} 
\label{sec:analysis}

We consider the performance of the proposed signals when transmitted over the AWGN channel.    We will consider the asymptotic performance for high values of $E_b/N_0$, $E_b$ being the mean energy per information bit. No channel coding is assumed to be used, thus $E_b=E_S$, Let us denote by ${\bm e}={\bm a}-\hat{\bm a}$ the sequence representing the difference between the transmitted sequence ${\bm a}$ and the erroneous one $\hat{\bm a}$. Without loss of generality, we will assume that any considered error event starts at time $n=0$.
We will also denote the normalized squared Euclidean distance~\cite{AnAuSu86}
\begin{equation}
d^2({\bm e}) = \frac{1}{E_b} ||s(t,{\bm a}) - s(t,\hat{{\bm a}}) ||^2\,. \label{eq:nor_dis}
\end{equation}

The probability of bit error for the optimal MAP sequence detector (implemented through the Viterbi algorithm) is well approximated by~\cite{PeRi05}
\begin{equation}\label{eq:prob_error}
P_b \approx \frac{  n_{ {\bm e}_{min} }   m_{ {\bm e}_{min} }    }{2^{ R_{ {\bm e}_{min} }  }}  {\rm Q}(\sqrt{d_{min}^2 E_b/N_0})
\end{equation}
where 
\begin{align}
d_{min}&=\min_{\bm e} d({\bm e})\,.\\
{\bm e}_{min}&=\arg\!\min_{\bm e} d({\bm e})\,,
\end{align}
$n_{ {\bm e}_{min}}$ is the number of bit errors (i.e., on the sequence $\{b_n\}$) caused by the error event ${\bm e}_{min}$, $m_{{\bm e}_{min}} = 2 \prod_{i=0}^{R_{{\bm e}_{min}}-1} (2-\frac{|e_{min,i}|}{2})$,
 $R_{{\bm e}_{min}}$ is the span of
symbol times where ${\bm e}_{min}$ is different from zero and ${\rm Q}(x)$ is the Gaussian Q function. If there are more sequences ${\bm e}$ corresponding to  $d_{min}$, the bit error probability will have more terms of the form of the right hand side of (\ref{eq:prob_error}), each one corresponding to a different sequence 
${\bm e}$. The coefficient $\frac{  n_{ {\bm e}_{min} }   m_{ {\bm e}_{min} }    }{2^{ R_{ {\bm e}_{min} }  }}$ is often called \textit{multiplicity} of the error event with minimum distance~\cite{BeBi99}.

Now, it only remains to identify the error events corresponding to $d_{min}$. This can be done by working on the 
phase tree, as described in~\cite{AnAuSu86}. 
%Let us consider the phase of the CPM signal in (\ref{e:bdb_sig}), denoted in the following as $\phi(t,{\bm a})$. This tree is formed by the ensemble of phase trajectories having a common start phase, say zero, at time $t=0$. We assume that data symbols for all the phase trajectories in the tree before this time are equal (i.e., we are considering an error event starting at time $t=0$). The phase trajectories do not coincide over the first symbol interval $(e_0 \neq 0)$. However, when going further into the tree, it is possible to find a pairs of phase trajectories which coincide (modulo $2 \pi$) at a specific time. This is referred to as a \textit{merger}. Let $t_m = mT$ denote the time where the two phase trajectories merge. The calculation of the Euclidean distance between both signals can be reduced to the interval $[0, t_m]$ The mergers are easily identified in the phase difference tree, since they correspond to a phase difference trajectory which is identically equal to zero for all $t \geq t_m$.  The mergers can be classified by sorting the values of $t_m$ in ascending order. The $1^{st}$ order mergers are the ones having the smallest $t_m$. Let ${\bm a}$ and $\hat{{\bm a}}$ be one pair of sequences yielding a phase difference trajectory merging at time $t_m$.  These two paths remain identical if $\phi(t,{\bm a})=\phi(t,\hat{{\bm a}})$, apart from a multiple of $2\pi$. The minimum distance can be found by using only the mergers of first orders which usually occur at time $(L+1)T$~\cite{AnAuSu86}.
We considered different modulation formats and computed the corresponding parameters $n_{ {\bm e}_{min} }$, $m_{ {\bm e}_{min} }$,  $R_{ {\bm e}_{min} }$, and $d_{min}$, for both the classical binary schemes and the proposed ones. We considered a modulation format with a rectangular phase pulse of length $L=1$ (1REC) and $h=1/2$, a modulation format with raised cosine frequency pulse of length $L=2$ (2RC) and $h=1/4$, and a Gaussian frequency pulse with normalized 3-dB bandwidth $\beta=0.5$ truncated to a length $L=2$ (2GAU) and $h=1/3$~\cite{AnAuSu86}.
The results are reported in Table~\ref{tab1}. It can be observed that the proposed schemes do not modify the minimum distance and, at most, they modify the multiplicity of the error event by a factor 2, 
although we do not have a formal proof that this always holds. In particular, the 2nd proposed precoder provides the same asymptotic performance as the classical scheme except for $h=1/2$, when the multiplicity is halved. 

\begin{table*}[!t]
\centering \protect\caption{Parameters for the computation of the asymptotic bit error probability
for three different phase pulses and modulation indices.}

\label{tab1} %
\begin{tabular}{|l|l|l|l|l|l|l|}
\hline 
\multicolumn{2}{|l|}{} & ${\bm e}_{min}$  & $n_{{\bm e}_{min}}$  & $m_{{\bm e}_{min}}$  & $d_{min}^{2}$  & $P_{b}$ \tabularnewline
\hline 
\multirow{3}{*}{%
\begin{tabular}{@{}l@{}}
1REC\tabularnewline
$h=1/2$\tabularnewline
\end{tabular}} & classical  & $(2,-2)$,$(2,2)$  & 2  & 4  & 2  & $2{\rm Q}\left(\sqrt{2\frac{E_{b}}{N_{0}}}\right)$ \tabularnewline
\cline{2-7} 
 & AMI precod.  & $(2,-2)$  & 2  & 4  & 2  & $2{\rm Q}\left(\sqrt{2\frac{E_{b}}{N_{0}}}\right)$ \tabularnewline
\cline{2-7} 
 & 2nd precod.  & $(2,-2)$  & 2  & 2  & 2  & ${\rm Q}\left(\sqrt{2\frac{E_{b}}{N_{0}}}\right)$ \tabularnewline
\hline 
\multirow{3}{*}{%
\begin{tabular}{@{}l@{}}
2RC\tabularnewline
$h=1/4$\tabularnewline
\end{tabular}} & classical  & $(2,0,-2)$  & 2  & 4  & 0.66  & ${\rm Q}\left(\sqrt{0.66\frac{E_{b}}{N_{0}}}\right)$ \tabularnewline
\cline{2-7} 
 & AMI precod.  & $(2,0,-2)$  & 2  & 8  & 0.66  & $2{\rm Q}\left(\sqrt{0.66\frac{E_{b}}{N_{0}}}\right)$ \tabularnewline
\cline{2-7} 
 & 2nd precod.  & $(2,0,-2)$  & 2  & 4  & 0.66  & ${\rm Q}\left(\sqrt{0.66\frac{E_{b}}{N_{0}}}\right)$ \tabularnewline
\hline 
\multirow{3}{*}{%
\begin{tabular}{@{}l@{}}
2GAU\tabularnewline
$\beta=0.5$\tabularnewline
$h=1/3$\tabularnewline
\end{tabular} } & classical  & $(2,0,-2)$  & 2  & 4  & 1.06  & ${\rm Q}\left(\sqrt{1.06\frac{E_{b}}{N_{0}}}\right)$ \tabularnewline
\cline{2-7} 
 & AMI precod.  & $(2,0,-2)$  & 2  & 8  & 1.06  & $2{\rm Q}\left(\sqrt{1.06\frac{E_{b}}{N_{0}}}\right)$ \tabularnewline
\cline{2-7} 
 & 2nd precod.  & $(2,0,-2)$  & 2  & 4  & 1.06  & ${\rm Q}\left(\sqrt{1.06\frac{E_{b}}{N_{0}}}\right)$ \tabularnewline
\hline 
\end{tabular}
\end{table*}

\section{Simulation Results} 
\label{sec:sim_res}

We first motivate the adoption of the proposed schemes by assessing their robustness in the case of a significant modulation index mismatch. We consider uncoded transmissions, both classical and proposed schemes, and the performance of the Viterbi-based receiver working on the principal components of the Laurent decomposition and designed for the nominal modulation index. In all cases considered in this paper, the performance of the suboptimal receivers based on the principal components of the Laurent decomposition is practically indistinguishable from that of the optimal detectors.

\begin{figure}[!t]
\center
\includegraphics[width=9.3cm]{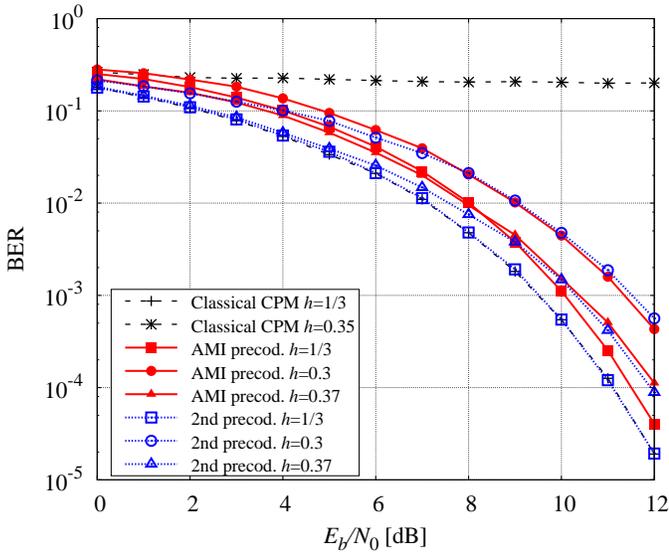}
\vspace{-7mm}
\caption{Robustness for the case of a 2GAU frequency pulse with $\beta=0.5$.}
\label{fig_unkown}
\end{figure}

As a first example, we consider a case inspired by the Bluetooth standard, where the 2GAU frequency pulse with $\beta=0.5$  is employed (third row of Table~\ref{tab1}). We will consider both classical and proposed schemes. We will assume that the modulation index at the transmitter can vary from 0.3 to 0.37 and that the receiver is designed for a nominal value of $h_{rx}=1/3$. No attempt is made at the receiver to compensate for the modulation index mismatch. In Fig.~\ref{fig_unkown}, we show the performance that is obtained when at the transmitter the nominal value or the values at the boundaries of the range is employed (at least for the proposed schemes, whereas for the case of a classical CPM a much lower mismatch is considered). It can be observed that, whereas for the classical binary CPM format a small value of the mismatch produces a large degradation, for both proposed schemes the degradation is very limited (at most 2 dB) in the considered range. 

\begin{figure}[!t]
\center
\includegraphics[width=9.3cm]{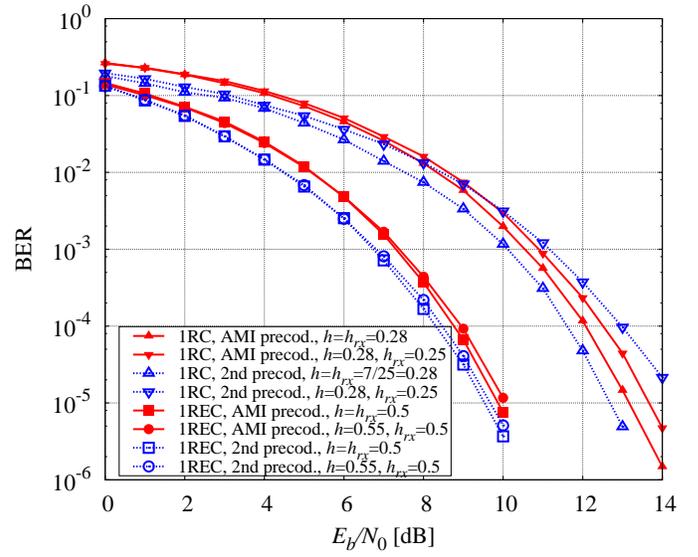}
\vspace{-7mm}
\caption{Robustness to a modulation index mismatch for 1REC and 1RC frequency pulses.}
\label{fig_unkown_BT}
\end{figure}

Similarly, in Fig.~\ref{fig_unkown_BT} we consider the case of full response CPMs with REC and RC frequency pulses. We remember that RC pulses attain a higher spectral compactness at the cost of a lower Euclidean distance. This can be also observed from the values of the 99.9\%-bandwidth reported in Table~\ref{tab_band}. In this case, the modulation index employed at the transmitter ($h$) and that at used to design the receiver ($h_{rx}$) are explicitly reported in the caption.
The results confirm our intuition about the robustness of the proposed schemes that exhibit a negligible performance loss even when the mismatch is large.

\begin{figure}[!t]
\center
\includegraphics[width=9.3cm]{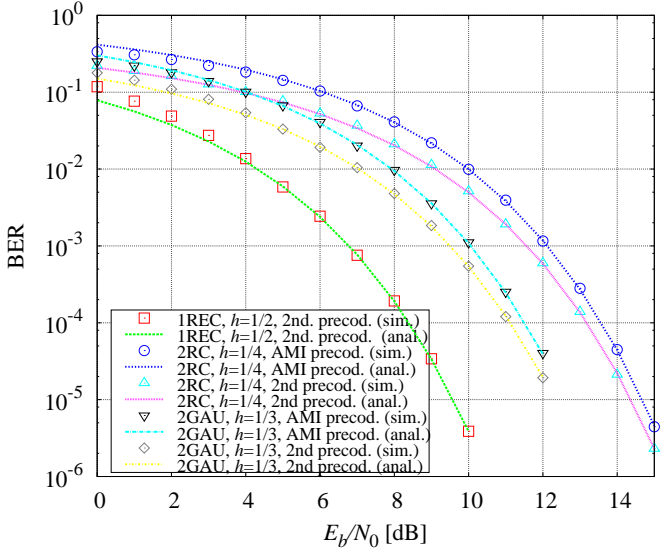}
\vspace{-7mm}
\caption{BER performance in the case of absence of modulation index mismatch (simulations  and closed-form asymptotic expressions).}
\label{fig:h_known}
\end{figure}

The correctness of the proposed asymptotic analysis for uncoded transmissions is addressed in Fig.~\ref{fig:h_known}, where the bit error rate (BER) performance, computed through simulations, for some of the modulation formats also considered in Table~\ref{tab1} is shown and compared with the asymptotic formulas reported in the table. There is no modulation index mismatch.  and the receivers considered in the simulations are not the optimal ones but the simplified receivers based on the Laurent decomposition. The fact that performance simulations using the suboptimal receiver is in perfect agreement with the analysis, which has been carried out with reference to the optimal detector, is a further evidence of what stated before, i.e., that the performance of the suboptimal receivers based on the principal components of the Laurent representation is practically indistinguishable from that of the optimal detectors.
 
\begin{figure}[!t]
\center
\includegraphics[width=9.3cm]{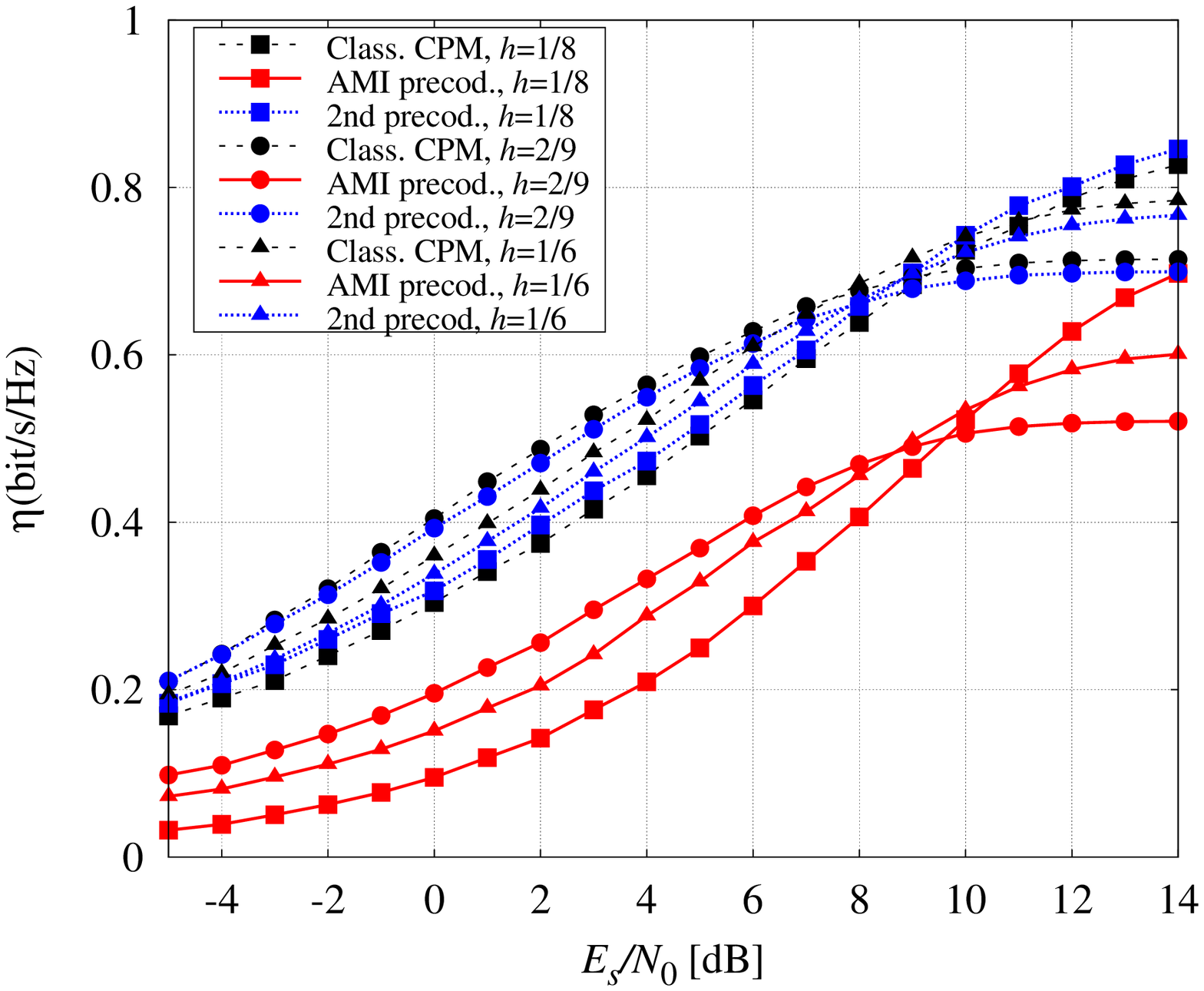}
\vspace{-7mm}
\caption{Spectral efficiency for 1REC modulations with different modulation indexes.}
\label{fig:spe_1REC}
\end{figure}

\begin{figure}[!t]
\center
\includegraphics[width=9.3cm]{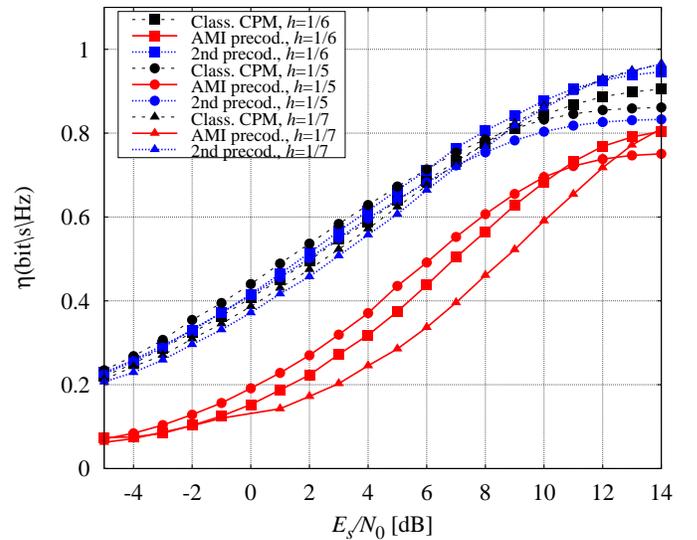}
\vspace{-7mm}
\caption{Spectral efficiency for 2RC modulations with different modulation indexes.}
\label{fig:spe_2RC}
\end{figure}

We now consider coded transmissions. In order to assess the ultimate performance limits of coded transmissions, we consider the spectral efficiency $\eta$ as defined in (\ref{eq:se}) instead of the simple information rate, in order to also capture  a possible bandwidth expansion caused by the considered precoders. In Figs.~\ref{fig:spe_1REC} and \ref{fig:spe_2RC}  we report the results in the case of a 1REC or a 2RC frequency pulse, respectively, with different modulation indices. The spectral efficiency will be shown as a function of $E_S/N_0$, which is related to $E_b/N_0$ by
$$
\frac{E_S}{N_0}=I\frac{E_b}{N_0}=\eta BT\frac{E_b}{N_0}\,.
$$
The second proposed precoder has a spectral efficiency quite similar to that of classical binary CPMs. On the contrary, a significant spectral efficiency degradation is observed when using the AMI precoder. This is due to a bandwidth expansion related to the use of this precoder which is related to the high occurrence of the phase transitions generated by the alternation $2 \rightarrow -2$ and vice versa.
This can be observed by looking at Table~\ref{tab_band}, where the values of $99.9\%$-bandwidth (defined as the bandwidth that contains $99.9\%$ of the overall signal power) is reported for the modulation formats considered in Figs.~\ref{fig_unkown_BT} and \ref{fig:h_known}. If, from one side, the use of the second proposed precoder does not entail significant modifications in the bandwidth values with respect to the classical binary CPM with the same modulation index and frequency pulse, a significant bandwidth expansion is observed with the AMI precoder. 

\begin{table}[h]
\centering
\caption{$99.9\%$-bandwidth for some of the considered schemes.}
\begin{tabular}{|l|l|l|l|l|l|l|l|l|}
\hline
                     & \multicolumn{4}{l|}{$B_{99.9\%}T$ for 1REC} & \multicolumn{4}{l|}{$B_{99.9\%}T$ for 2RC} \\ \hline
$h$ & 1/7       & 1/6       & 2/9      & 1/8      & 2/7       & 1/6      & 1/5      & 1/7      \\ \hline
Classical           & 1.21      & 1.27      & 1.4      & 1.18     & 1.24      & 1.1      & 1.16     & 1.02     \\ \hline
AMI prec.        & 1.51      & 1.8       & 2.2      & 1.39     & 1.76      & 1.32     & 1.43     & 1.24     \\ \hline
2nd prec.    & 1.22      & 1.3       & 1.43     & 1.03     & 1.4       & 1.03     & 1.2      & 1.03     \\ \hline
\end{tabular}
\label{tab_band}
\end{table}

\begin{figure}[!t]
\center
  \includegraphics[width=9cm]{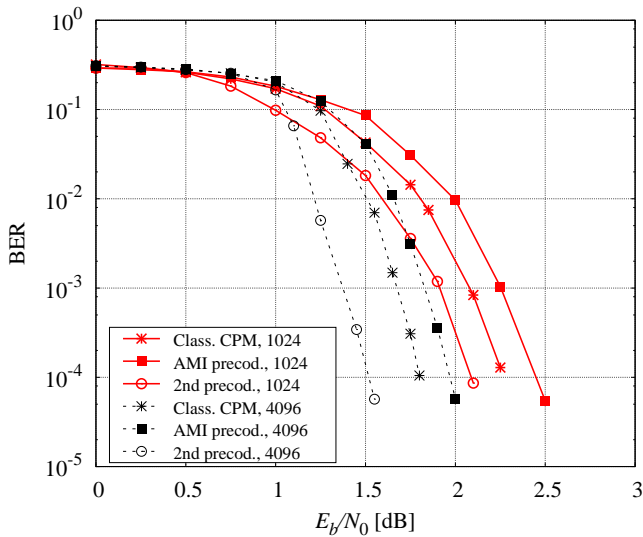}
\caption{BER performance for a 1REC modulation with $h=1/2$ serially concatenated with a convolutional encoder via a pseudo-random interleaver.}
\label{fig:sccpm}
\end{figure}

Finally, in Fig.~\ref{fig:sccpm} we considered BER simulations for a coded transmission system. We serially concatenated the proposed schemes with a binary convolutional encoder with generators $(7, 5)$ (octal notation) through a pseudo-random interleaver of length 1024 or 4096 bits. We used a 1REC frequency pulse with modulation index $h=1/2$. We also report the performance related to the use, in the same concatenation, of a classical CPM scheme with the same frequency pulse and modulation index, corresponding to a minimum shift keying (MSK) modulation.  For all considered schemes, a number of 16 iterations between detector and decoder is allowed. This figure demonstrates that the proposed schemes are suitable for such kind of concatenation and an interleaver gain can be observed due to the preserved recursive nature of the modulator.

\section{Conclusions}
\label{se:concl}
We proposed new binary CPM schemes for applications where low-cost transmitters are employed. In these applications, the modulation index of the transmitter can be much more different than the nominal one and, if not properly taken into account at the receiver through the use of techniques for estimation and compensation, this mismatch can severely degrade the performance. 
The proposed new schemes are based on the  concatenation of a precoder with binary input and ternary output, and a ternary CPM scheme. The aim of the precoder is to constrain the evolution of the CPM phase state and this makes them more robust to a modulation index mismatch. Two precoders have been described and investigated in this paper. 
The overall scheme resulting from the adoption of the first precoder has only two states, independently of the adopted modulation index but the performance, in terms of spectral efficiency is worse than that of classical CPM due to a bandwidth expansion. The second precoder generates an overall scheme with a number of states which depends on the modulation index denominator.
We studied these schemes from the point of view of coded and uncoded performance and spectral characteristics, demonstrating that they can be a valid solution in these low-cost applications.

\end{document}